\pdfoutput=1
 \documentclass{article}
 \usepackage{setspace}   
\usepackage{dcolumn}
\usepackage{sectsty}
\sectionfont{\centering}
\usepackage[utf8]{inputenc}
\usepackage[margin=1in]{geometry}
   
 \usepackage{hyperref}
 \hypersetup{
    pdftoolbar=true,        
    pdfmenubar=true,      
    pdffitwindow=false,    
    pdfstartview={FitH},
    pdfauthor={Sabina Yeasmin},    
    colorlinks=true,      
    linkcolor=blue,        
    citecolor=blue,      
}

 \usepackage{amsmath}
 \usepackage{amssymb}
\usepackage{breqn}
 \usepackage[backend=biber,sorting=none]{biblatex}
\usepackage{graphicx}
\begin{document}
  
   \begin{center}
       \vspace*{1cm}

       \text{\Large Effect of Dissipation on Warm Chromo-Natural Inflation}

       \vspace{0.5cm}
        Sabina Yeasmin *, Atri Deshamukhya\\
            
       \vspace{0.25cm}

       \textit{Department of Physics, Assam University, Silchar, Assam 788011, India }\\
       \end{center} 
       
       \vspace{0.5cm}
       \begin{center}
  \begin{minipage}{0.85\textwidth}
  
{ {\small We examined the chromo-natural inflation  in the context of warm inflation with variable dissipation coefficient. The dynamical equations of this model are obtained. We studied the cosmological perturbation theory in this model. The sources of density fluctuations in this model are mainly the thermal fluctuations of the inflaton field like general warm inflationary model. Finally, cosmological observables, namely, the spectral index  and tensor to scalar ratio are calculated. It is found that the cosmological observables are consistent with observational data and the tensor to scalar ratio is smaller than that in the chromo-natural inflation.}  }
    
    \end{minipage}        
    \end{center} 
    \vspace{0.5cm}
 \begin{center}
     \section{\label{sec:levelI}Introduction}
 \end{center}

The idea of inflation was first proposed by Alan Guth \cite{PhysRevD.23.347} in 1981 to solve the problems in the big bang model, notably the horizon, flatness and horizon  problem and further developed by Linde \cite{1982PhLB..108..389L}\cite{LINDE1983177}\cite{linde2005particle}\cite{PhysRevD.49.748}, Albrecht \cite{PhysRevLett.48.1220} and many others. It is the most successful model so far for explaining the large scale structure of the universe. Inflationary dynamics can be  realized in two different ways, one is cold inflation and another is warm inflation. In standard cold scenario, the scalar field named inflaton is assumed to have no interaction with the other fields those may be present in the universe. So, there is no possibility for radiation to be produced during the process, thus leading to a thermodynamically supercooled phase of the Universe at the end of inflation. So, a reheating period is needed to be invoked at the end of inflation to fill the universe with radiation and make it enter the radiation dominated phase. In an alternate picture, termed warm inflation \cite{PhysRevLett.75.3218}\cite{PhysRevD.55.3346}\cite{54db6c2c12e0450d9e603224eb51880c}\cite{PhysRevD.58.123508}\cite{DESHAMUKHYA_2009}, the inflaton field is assumed to be coupled to several other fields during the inflationary period. The inflaton dissipates its energy into radiation during inflationary process. So, in warm inflation radiation production can simultaneously occur with the expansion of the universe and there is a smooth transition from the inflationary phase to radiation dominated phase without the necessity of introducing a separate reheating phase. The temperature of the thermal bath should be greater than the Hubble parameter i.e. $(T >H)$ in warm inflation \cite{Berera_2009}. The dissipative coefficient $\Gamma$ plays a crucial role in describing the dynamics of warm inflation. Depending on the strength of dissipation, there are two dissipative regimes in warm inflation, strong and weak. In strong dissipative regime, the dissipation coefficient is greater than the Hubble parameter and in the weak dissipative regime, the dissipation coefficient is smaller than the Hubble parameter \cite{Bastero_Gil_2005}
 
 A successful inflationary model must provide sufficiently long period of inflation to solve the horizon problem. To satisfy these constraints, the potential for the inflaton must be very flat. Natural inflation\cite{PhysRevLett.65.3233}\cite{Adams_1993}\cite{Savage_2006}\cite{2015}\cite{2004} is one of the viable models in which shift symmetry assures a flat inflaton potential. But Natural inflation requires a large axion decay constant, $f\sim M_P$ to match the constraints on the spectral index, which appears to be difficult to realize within the framework of string theory\cite{Banks_2003}. Later in 2012, Adshead et al \cite{2012} proposed an axionic inflationary model which has a sub-Planckian axionic decay constant. There in addition to the axionic field, they considered a collection of non-abelian gauge fields having a rotationally invariant vacuum expectation value. This model is known as Chromo-Natural Inflation \cite{2012}\cite{Adshead_2013}\cite{Dimastrogiovanni_2013}\cite{Bhattacharjee_2015}. In this paper, we have extended Chromo-Natural inflation to a warm inflationary scenario, termed 'Warm Chromo-Natural Inflation'.

 \vspace{8pt}
 \hrule
 \vspace{5pt}
 { \footnotesize *E-mail: sabina.yeasmin@aus.ac.in}
 
Studying  Chromo-natural inflation in the context of warm inflation is more natural and proper. Because the coupling between the axion and the non-abelian gauge field can cause the production of thermal bath during inflation period which has been neglected in chromo-natural inflation. In warm chromo-natural inflation, the thermal bath produced as a result of the dissipation of the inflaton's energy through its coupling with the other field is considered and plays an important role in its dynamics. In this work, we have studied the dissipative effects that modify the dynamics of warm chromo-natural inflation. We consider three forms of dissipation coefficient, (i) a constant dissipation coefficient, (ii) a dissipation coefficient linearly dependent on the temperature of the thermal bath $(\Gamma \propto T)$, and (iii) a dissipation coefficient dependent on the inflaton and temperature of the thermal bath $(\Gamma \propto \frac{T^3}{\chi^2})$. We are interested to analyse the warm chromo-natural inflation model in both the strong and weak dissipative regimes .


This paper is organized as follows: In section 2, we briefly introduce the basic dynamics of the inflaton field in warm chromo-natural inflation and discuss the spectra of scalar and tensor perturbations which are different from that in chromo-natural inflation. In section 3.1, we investigate the effect of the three forms of dissipation coefficients in the weak dissipative regime and estimate the model parameters by virtue of Planck results. The effect of the three forms of dissipation coefficients in the strong dissipative regime are studied and best fitted parameters are computed in section 3.2. And finally in section 4, we draw the conclusion.
 \vspace{.5cm}

\section{ Dynamics of the Warm Chromo-Natural Inflation}

  We consider the action that contains the gauge field and the pseudo scalar axion field: \cite{2012}\cite{PhysRevD.87.103501}
\begin{eqnarray}
    S_{chromo}&=&\int d^4x\sqrt{-g}\left[-\frac{M_P^2}{2}R-\frac{1}{4}F_{\mu\nu}^a F^{\mu\nu}_a-\frac{1}{2}(\partial\chi)^2-\mu^4\left(1+\cos\left(\frac{\chi}{f}\right)\right)\nonumber\right.\\&+&\left.\frac{\lambda}{8f}\chi\epsilon^{\mu\nu\rho\sigma}F_{\mu\nu}^a F^a_{\rho\sigma}\right]
\end{eqnarray}
Where $ F_{\mu\nu}^a =\partial_\mu A_\nu^a-\partial_\nu A^a_\mu-\bar g f^{abc} A_\mu^b A_\nu^c$, $\chi$ and $A^a_\mu $ are the axion and the $ SU(2)$ gauge fields respectively. Here, $\hbar=c=M_P=1 $ (in natural units). The last term in the action is the interaction term (Chern-Simon term) which describes the interaction between the axionic field and the gauge field. Here,  $ \bar g $ is the gauge coupling constant, $ f $ is the axion decay constant, $\mu$ is the mass scale of the theory, $\lambda$ is the gauge parameter and $f_{abc}$ is the structure constant of the gauge group with normalization condition $f^{123}=1$. The background metric is the standard FRW metric, $ds^2=-dt^2+a^2(t)\delta_{ij}dx^idx^j$

  When the mass of the fluctuation is much heavier than the Hubble scale, the linear fluctuation of the gauge field which introduces a perturbative instability will disappear. Under this condition gauge field can be integrated out in a single field effective action \cite{Dimastrogiovanni_2013} which involves only the axion field taking the form :
  \begin{equation}
    S=\int d^4x\sqrt{-g}\left[-\frac{M_P^2}{2}R -\frac{1}{2}(\partial\chi)^2+\frac{1}{4 \Lambda^4}(\partial\chi)^4-\mu^4\left(1+\cos\left(\frac{\chi}{f}\right)\right) \right]
\end{equation}
The term $\frac{1}{4 \Lambda^4}(\partial\chi)^4 $ captures the effect of the gauge field in the single field effective theory with, $\Lambda^4=\frac{8f^4\bar g^2}{\lambda^4}$\\
The above effective action can be expressed as:

\begin{equation}
    S =\int d^4x\sqrt{-g}\left[-\frac{M_P^2}{2}R +P(X,\chi)\right]
\end{equation}
Where $P(X,\chi)$ is the matter Lagrangian density and has the expression 
 
\begin{equation}
    P(X,\chi)=X+\frac{X^2}{\Lambda^4}-V(\chi)
    \label{a}
\end{equation}
with $ X=-\frac{1}{2}(\partial\chi)^2$  
and $V(\chi)= \mu^4\left(1+\cos\left(\frac{\chi}{f}\right)\right)$ is the potential of the axionic field $\chi$.\\
 The energy-momentum tensor of the inflaton field is
\begin{equation}
  T_{\mu\nu}=(\rho_\chi+p_\chi)u_\mu u_\nu+p_\chi g_{\mu\nu}  
  \label{b}
\end{equation}
where $\rho_\chi$, $ p_\chi$ and $u$ are the energy density, pressure and 4-velocity of the inflaton field respectively.
But, the energy-momentum tensor of the inflaton field can be obtained by varying the Lagrangian of the inflaton field with respect to metric and can be expressed as 
\begin{equation}
    T_{\mu\nu}=P_{,X}\partial_\mu\chi \partial_\nu\chi+g_{\mu\nu}P(X,\chi)
    \label{c}
\end{equation}
where $P_{,X}$ denotes derivative of $P$ with respect to $X$. Comparing equation \eqref{a},\eqref{b} and \eqref{c} we find 
\begin{equation}
    \rho_\chi=X+\frac{3X^2}{\Lambda^4}+V(\chi)
    \label{d}
\end{equation}
\begin{equation}
 p_\chi=X+\frac{X^2}{\Lambda^4}-V(\chi)
 \label{e}
 \end{equation}
 Here pressure term $p_\chi$ is same as the matter Lagrangian density $P$.
Now, the basic dynamical equations in the warm inflationary scenario are \cite{2016}\cite{2018}\cite{Seery_2005},  
\begin{equation}
    H^2=\frac{\rho}{3M_P^2}
    \label{100}
\end{equation}
 
\begin{equation}
    \dot{\rho}_\chi+3H(\rho_\chi+P)=-\Gamma \dot{\chi}^2
    \label{f}
\end{equation}

\begin{equation}
    {\dot{\rho}_\gamma}+4H\rho_\gamma=\Gamma\dot{\chi}^2
    \label{101}
\end{equation}
 Where $ \Gamma $ is the dissipation coefficient which describes the decay of the inflaton field into radiation during the process of inflation. Now, considering a homogeneous inflaton field, $ X=\frac{\dot{\chi}^2}{2} $, the equation of motion can be obtained by substituting equations \eqref{d} and \eqref{e}, in equation \eqref{f} as,
 \begin{equation}
    \ddot{\chi}\left(1+\frac{3\dot{\chi}^2}{\Lambda^4}\right)+3H\dot{\chi}\left(1+\frac{\dot{\chi}^2}{\Lambda^4}\right) +\Gamma\dot{\chi}=\frac{\mu^4}{f}\sin\left(\frac{\chi}{f}\right)
\end{equation}
or \begin{equation}
    \ddot{\chi}\left(1+\frac{3\dot{\chi}^2}{\Lambda^4}\right)+3H\dot{\chi}\left(1+\frac{\dot{\chi}^2}{\Lambda^4}  + r\right)=\frac{\mu^4}{f}\sin\left(\frac{\chi}{f}\right)
    \label{102}
\end{equation}
Where the dissipation rate is defined as $r=\frac{\Gamma}{3H}$.

We can have different scenarios, depending on the ratio $r$. When $r<1$, the effect of dissipation is weak. This is called weak dissipative warm inflation. And when $r>1$, the scenario is called strong dissipative warm inflation.
 Now, for inflation to occur, the inflaton process must be potential energy dominated (slow-roll approximation). i.e. in the slow-roll regime $\rho\simeq V(\chi)$, the dynamical equations \eqref{100},\eqref{102} and \eqref{101} can be written as
 
 \begin{equation}
    H^2=\frac{V}{3M_P^2}
    \label{103}
\end{equation}
 
\begin{equation}
    3H\dot{\chi}\left(1+\frac{\dot{\chi}^2}{\Lambda^4}  + r\right)=\frac{\mu^4}{f}\sin\left(\frac{\chi}{f}\right)
    \label{104}
\end{equation}
and
\begin{equation}
    \rho_\gamma=\frac{\Gamma\dot{\chi}^2} {4H}=C T^4
    \label{105}
\end{equation}
 where $T$ is the temperature of the thermal bath and $C=\frac{g_* \pi^2}{30}$ is the Stefan-Boltzmann constant ($g_*$ represents the number of degrees of freedom of radiation field).
 The consistency of the slow-roll approximation is governed by a set of slow-roll parameters. These slow-roll parameters $ \epsilon$, $\eta$ and $\beta$ are defined as
\begin{equation}
    \epsilon=-\frac{\dot{H}}{H^2}
    \label{106}
\end{equation}
\begin{equation}
    \eta=-\frac{\ddot{H}}{2H\dot{H}}
    \label{107}
\end{equation}
\begin{equation}
    \beta=-\frac{\dot{\rho}_\gamma}{H\rho_\gamma}
    \label{108}
\end{equation}
  The slow-roll parameter $\beta$ is additional to the warm inflationary scenario which takes into account the variation of dissipation coefficient $\Gamma$.
 In terms of potential $ V(\chi) $ of the inflaton field, these slow-roll parameters can be  expressed as
 \begin{equation}
    \epsilon=\frac{M_P^2}{2\left(1+\frac{\dot{\chi}^2}{\Lambda^4}+r\right)}\left(\frac{V^\prime}{V}\right)^2
    \label{109}
\end{equation}
\begin{equation}
    \eta=\frac{M_P^2}{\left(1+\frac{\dot{\chi}^2}{\Lambda^4}+r\right)}\left(\frac{V^{\prime\prime}}{V}\right)
    \label{110}
\end{equation}
\begin{equation}
    \beta=\frac{M_P^2}{\left(1+\frac{\dot{\chi}^2}{\Lambda^4}+r\right)}\left(\frac{\Gamma^\prime V^\prime}{\Gamma V}\right)
    \label{111}
\end{equation}
For slow-rolling of inflation field, the slow-roll parameters have to satisfy the conditions \cite{Hall_2004}\cite{Moss_2008}\cite{PhysRevD.62.083517}, $\epsilon<<1$, $|\eta|<<1$ and $|\beta|<<1$ during inflationary phase.

The number of e-foldings when the inflation field $\chi$ rolls from its value $\chi_i$ to $\chi_f$ is estimated as:
\begin{equation}
    N=\int^{t_2}_{t_1} H dt=\int^{\chi_f}_{\chi_i}\frac{H}{\dot{\chi}}d\chi
    \label{115}
\end{equation}

\subsection{Calculation of Perturbation Spectra}

   We will consider the thermal fluctuations of the inflaton field in the spatially flat gauge and assume that $\Gamma=\Gamma(\chi)$. The perturbed FRW metric in this gauge \cite{2018} is 
  \begin{equation}
      ds^2=-(1+2A)dt^2+2a(t)\partial_iBdtdx^i+a^2(t)\delta_{ij}dx^idx^j
\end{equation}
  Where $A(x,t)$ and $B(x,t)$ are scalar perturbations of the metric. Now, the linear perturbation in the energy density $(\delta\rho)$ and the pressure $(\delta p)$  and the scalar part of the linear perturbation of the 4- velocity $(\delta u)$ in the flat gauge can be expressed as
  \begin{equation}
    \delta \rho= \dot{\chi} \dot{\delta \chi}+\frac{3}{\Lambda^4}\dot{\chi}^3  \dot{\delta\chi}-A\dot{\chi}^2-\frac{3}{\Lambda^4}A\dot{\chi}^4+V_\chi \delta \chi
    \label{1}
\end{equation}
\begin{equation}
    \delta u=- \frac{\delta \chi}{\dot{\chi}} 
    \label{2}
\end{equation}
\begin{equation}
    \delta p=\dot{\chi}\dot{\delta\chi}+\frac{1}{\Lambda^4}\dot{\chi}^3  \dot{\delta\chi}-A\dot{\chi}^2-\frac{1}{\Lambda^4}A\dot{\chi}^4-V_\chi \delta \chi
    \label{3}
\end{equation}

   Using equations \eqref{1}-\eqref{3}, we can derive the perturbed equation of the inflaton field in the momentum space. Thus the perturbed equation of inflaton field in the spatially flat gauge becomes  
  \begin{eqnarray}
  && \left(1+\frac{3}{\Lambda^4}\dot{\chi}^2\right)\ddot{\delta\chi}_k+\left[3H\left(1+\frac{3}{\Lambda^4}\dot{\chi}^2\right)+\Gamma\right]\dot{\delta\chi}_k+\left[ \frac{k^2}{a^2}\left(\frac{\dot{\chi}^2}{\Lambda^4}+1\right)\right]\delta\chi_k\nonumber\\&=&\left(1+\frac{3}{\Lambda^4}\dot{\chi}^2\right)\dot{\chi}\dot{A} 
  +\left(6H\dot{\chi}^3-\Gamma\dot{\chi}\right)A+\left[ \frac{k^2}{a}\left(\frac{\dot{\chi}^2}{\Lambda^4}+1\right)\right]\dot{\chi}B-\delta\Gamma\dot{\chi}-2A V_{,\chi}
  \label{152}
  \end{eqnarray}
  The metric terms on the right-hand-side obey the energy and momentum constraints
  \begin{equation}
      3H^2A+\frac{k^2}{a}HB=-4\pi G\delta\rho
      \label{150}
  \end{equation}
  \begin{equation}
      HA=4\pi G (\rho+p)\delta u 
      \label{151}
  \end{equation}
  The expressions of $A$ and $B$ from equations \eqref{150} and \eqref{151} are substituted in equation \eqref{152} to eliminate the metric perturbations. Introducing the thermal stochastic noise source $\xi_k(t)$ in the perturbed inflaton equation, we get
  \begin{equation}
     \left(1+\frac{3}{\Lambda^4}\dot{\chi}^2\right)\ddot{\delta\chi}_k(t)+\left[3H\left(1+\frac{3}{\Lambda^4}\dot{\chi}^2\right)+\Gamma\right]\dot{\delta\chi}_k(t)+\frac{k^2}{a^2}\left(\frac{\dot{\chi}^2}{\Lambda^4}+1\right)\delta\chi_k(t)=\xi_k(t)
     \label{12}
 \end{equation}
  
 Equation \eqref{12} is called the Langevin equation which describes the behavior of a scalar field interacting with radiation. Now, in the slow roll regime, the term $ \ddot{\delta\chi}_k$ can be neglected. Thus the Langevin equation becomes
 \begin{equation}
      3H\left(1+\frac{3}{\Lambda^4}\dot{\chi}^2+r\right)\dot{\delta\chi}_k(t)+\frac{k^2}{a^2}\left(\frac{\dot{\chi}^2}{\Lambda^4}+1\right)\delta\chi_k(t)=\xi_k(t)
     \label{13}
 \end{equation}
 The solution of the equation \eqref{13} is
  
 \begin{equation}
   \delta\chi_k(t)=\frac{1}{3H\left(1+\frac{3}{\Lambda^4}\dot{\chi}^2+r\right)}\exp{\left[-\frac{t}{\tau}\right]} \int_{t_0}^t\exp{\left[\frac{t^\prime}{\tau}\right]}\xi_k(t^\prime)dt^\prime+ \delta\chi_k(t_0)\exp{\left[-\frac{t-t_0}{\tau}\right]} 
 \end{equation}
 Where,
  $ \tau(\chi)=\frac{3H\left(1+\frac{3}{\Lambda^4}\dot{\chi}^2+r\right)}{\left(\frac{\dot{\chi}^2}{\Lambda^4}+1\right) \frac{k^2}{a^2}}=\frac{3H\left(1+\frac{3}{\Lambda^4}\dot{\chi}^2+r\right)}{\left(\frac{\dot{\chi}^2}{\Lambda^4}+1\right)k_p^2} $ and $k_p$ is the physical wave number. From the above equation, it is found that  when $k_p$ is larger, $\tau(\chi)$ will be smaller and as a result relaxation rate will be faster. If $k_p$  for one mode is sufficiently large  to relax within Hubble time, the mode will be thermal. When $k_p$ of one mode of $\delta\phi_k$  is smaller than the freeze-out physical wave number $k_F$,  the mode will not thermalize  during a Hubble time \cite{54db6c2c12e0450d9e603224eb51880c}. So, the freeze-out physical wave number $k_F$ can be obtained as    
 \begin{equation}
     k_F=\sqrt{3H^2\left( \frac{1+\frac{3}{\Lambda^4}\dot{\chi}^2+r}{\frac{\dot{\chi}^2}{\Lambda^4}+1}\right)}
     \label{14}
 \end{equation}
  In general, the thermal fluctuations of the inflaton field $\delta\chi_k$ in warm inflation can be expressed as \cite{54db6c2c12e0450d9e603224eb51880c}\cite{Berera_2009} 
  \begin{equation}
 \delta\chi_k^2=\frac{k_F T}{2\pi^2}
 \end{equation}
 From equation \eqref{14}, the thermal fluctuations of the inflaton field becomes
 \begin{equation}
 \delta\chi^2=\frac{H T}{2\pi^2} \sqrt{3\left( \frac{1+\frac{3}{\Lambda^4}\dot{\chi}^2+r}{\frac{\dot{\chi}^2}{\Lambda^4}+1}\right)}
 \end{equation}
 
\vspace{.5cm}

   The power spectrum for the scalar fluctuations \cite{PhysRevD.62.083517} in warm inflation is $P=\left(\frac{H}{\dot{\chi}}\right)^2\delta\chi^2$. Thus, the spectrum of scalar perturbations in the warm chromo-natural inflation :
\begin{eqnarray} 
  P_R&=&\frac{H^3 T}{4\pi^2 X}\sqrt{3\left( \frac{1+\frac{3}{\Lambda^4}\dot{\chi}^2+r}{\frac{\dot{\chi}^2}{\Lambda^4}+1}\right)}\\
  &=& \frac{T\left(\mu^4\left(1+\cos\left(\frac{\chi}{f}\right)\right)\right)^{3/2}}{2\sqrt{3}\pi^2\dot{\chi}^2M_P^3}\sqrt{1+\frac{\sqrt{3}\Gamma\Lambda^4M_P}{9\dot{\chi}^2\left(\mu^4 \left(1+\cos\left(\frac{\chi}{f}\right)\right)\right)^{1/2}}} 
  \label{117}
\end{eqnarray}
And the power spectrum for the tensor perturbation can be expressed as
    
\begin{eqnarray}
     P_T&=&\frac{2V}{3\pi^2M_P^4}\\
      &=&\frac{2\mu^4}{3\pi^2M_P^4} \left(1+\cos\left(\frac{\chi}{f}\right)\right)
      \label{118}
\end{eqnarray}
which has the same form as in cold inflation.
Now, the two important cosmological observables, namely, the spectral index $n_s$ and tensor to scalar ratio $R$ can be defined as
\begin{equation}
  n_s-1=\frac{d \ln P_R}{d \ln k}
  \label{124}
\end{equation}
\begin{equation}
    R=\frac{ P_T}{P_R}
    \label{125}
\end{equation}

 \subsection{Dissipation Coefficient}
  Dissipation coefficient was developed in super-symmetric models which have three superfields $X$, $Y$ and $Z$. $\chi$, $y$ and $z$ are the bosonic components for the  superfields $X$, $Y$ and $Z$ respectively, where $\chi$ describe the inflaton field. The inflaton field $\chi$ couples to bosonic component $y$ and fermionic component $\psi_y$ of the superfield $Y$ and gain masses $m_{\psi_y}=m_y=g\chi$ from the interaction with the inflaton $\chi$ which again transfer its energy to bosonic and fermionic components of the superfield $Z$. The field $Z$ thermalises and form the thermal bath. A general expression for the dissipation coefficient \cite{Zhang_2009} is given by 
  \begin{equation}
      \Gamma(\chi, T)=C_0 \frac{T^m}{\chi^{m-1}}
  \end{equation}
  where $C_0$ is connected to the dissipative microscopic dynamics. The form of dissipation coefficient depends on the dissipative channels of inflaton and the mass of the intermediate fields. In this paper, three forms of dissipation coefficient are considered, which are

  \begin{itemize}
      \item  Constant dissipation coefficient i.e.  
     \begin{equation}
      \Gamma=\Gamma_0    
      \end{equation}  
 \end{itemize}

 \begin{itemize}
      \item Temperature dependent dissipation coefficient,
      \begin{equation}
       \Gamma(T)=C_T  T  
      \end{equation}
        This form of dissipation coefficient arises in the high temperature limit, when the inflaton couples to the light bosonic and fermionic $(m_y<<T)$ intermediate fields.
  \end{itemize}
    
  \begin{itemize}
      \item And dependent on temperature and inflaton field,
      \begin{equation}
       \Gamma(\chi, T)=C_\chi \frac{T^3}{\chi^2}   
      \end{equation}
      The constant $C_\chi$ depends on the couplings and the multiplicities of the superfields. This form of dissipation coefficient arises in the low temperature limit, when the inflaton couples to the heavy bosonic $(m_y>>T)$ intermediate fields.
  \end{itemize}

  \vspace{.5cm}
   \section{Results}
  In this section, we numerically analyze the effect of the form of dissipation coefficient. For the purpose, we conduct the analysis in two dissipative regimes viz weak $r<<1$ and strong $r>>1$

  \subsection{Weak Dissipative Regime}
   
  In weak dissipative regime, $r<<1$. So, the slow-roll parameters $\epsilon$, $\eta$ and $\beta$ in this regime become,
  \begin{equation}
    \epsilon=\frac{M_P^2 \Lambda^4}{2 f^2 \dot{\chi}^2} \left(\frac{1-\cos{\frac{\chi}{f}}}{1+\cos{\frac{\chi}{f}}}\right) 
    \label{112}
  \end{equation}
  \begin{equation}
       \eta = -\frac{M_P^2 \Lambda^4}{f^2\dot{\chi}^2}  \left(\frac{\cos{\frac{\chi}{f}}}{1+\cos{\frac{\chi}{f}}}\right) 
       \label{113}
  \end{equation}
  \begin{equation}
   \beta= \frac{2 M_P^2 \Lambda^4}{ f \dot{\chi}^3}  \left(\frac{\sin{\frac{\chi}{f}}}{1+\cos{\frac{\chi}{f}}}\right)   
   \label{114}
  \end{equation}
 where we used $\frac{\dot{\chi}^2}{2}>>\Lambda^4$ (slow-roll hierarchy scaling limit). And the equation of motion of the inflaton field becomes
 \begin{equation}
    3H\dot{\chi}\left(\frac{\dot{\chi}^2}{\Lambda^4}  \right)=\frac{\mu^4}{f}\sin\left(\frac{\chi}{f}\right)
\end{equation}
 Using equation \eqref{103}, we get
 \begin{equation}
      \dot{\chi} = \frac{0.832683 M_P^{\frac{1}{3}} \mu^{\frac{2}{3}} \Lambda^{\frac{4}{3}} {\sin^{\frac{1}{3}} \left(\frac{\chi}{f}\right)}}{f^{\frac{1}{3}} \left(1+\cos \left(\frac{\chi}{f}\right)\right)^{\frac{1}{6}}}
      \label{128} 
 \end{equation}
 Substituting $\dot{\chi}$ in equations \eqref{112}, \eqref{113} and \eqref{114}, we get the expression of sow-roll parameters $\epsilon$, $\eta$ and $\beta$ in terms of model parameters. End of the inflation is governed by the violation of the slow-roll conditions. It is numerically checked that the slow-roll parameters  $\epsilon$ and $\eta$ violate the slow-roll condition i.e. reach unity at the same time and earlier than $\beta$. Here, $\epsilon=1$ is used to determine the final field value $\chi_f$ and
 initial field value $\chi_i $ can be obtained by using equation \eqref{115}. Therefore, $\chi_i $ is given by
 \begin{eqnarray}
      \chi_i=2f \sin^{-1} \left[\left[-\frac{M_P^{4/3}\Lambda^{4/3} N}{2^{1/3} 3^{2/3} f^{4/3} \mu^{4/3}}+\left(\sin\left({\frac{\chi_f}{2 f}}\right)\right)^{2/3}\right]^{3/2}\right]
      \label{127}
 \end{eqnarray}
 The largest density perturbations are obtained when $\chi=\chi_i$ \cite{PhysRevLett.65.3233} . So, the expressions for scalar power spectrum and tensor power spectrum at initial field value $(\chi_i)$ in weak dissipation regime become
    \begin{equation}
    P_R= \frac{0.0358588 r^{1/4}f^{1/2} \mu^5 \left(\cos \left(\frac{\chi_i}{f}\right)+1\right)^{7/4}}{M_P^{7/2} C^{1/4} \Lambda^2 \sin ^{\frac{1}{2}}\left(\frac{\chi_i}{f}\right)} 
    \label{121}
 \end{equation}
 \begin{equation}
    P_T=\frac{16\mu^4}{3\pi M_P^4}\left(1+\cos{\left(\frac{\chi_i}{f}\right)}\right)
    \label{122}
\end{equation}
\subsubsection{Case-1: $\Gamma=\Gamma_0$}
In this case, as we consider $\Gamma$ as constant, the slow-roll parameter $\beta$ is zero. Therefore, equation \eqref{127} holds in this case also and substituting $\chi_i$ in the expressions of $P_R$ and $P_T$, the power spectrum can be written in terms of model parameters $f$, $\mu$, $\Lambda$ and $r$ and using these, the spectral index $n_s$ and tensor to scalar ratio $R$ can be obtained in terms of model parameters. We take the values of the parameters that satisfy the observational constraints on $n_s$ and $R$ for 60 e-folds. The allowed values of $n_s$ and $R$ for different sets of model parameters are presented in the following table (Table 1). For different values of model parameters, we obtain the upper and lower limit of $r$ corresponding to values $0.5$ and $5\times10^{-6}$ respectively in this weak dissipative regime.

\vspace{.5cm}
\begin{center}
    Table 1:   The values of the spectral index $n_s$ and tensor to scalar ratio $R$ for different values of model parameters $ f$, $\mu$ and $\Lambda$ (in units of $M_P$) and $r$ with $N=60$
\end{center}
  \begin{center}
      \begin{tabular}{|c|c|c|c|c|c|c|}
      \hline
      $\mu$   & $ f $  & $ \Lambda$   & $ n_s$   & $ R$ & $r$\\ 
      
      \hline
      $ 0.0001$   & $0.005$   & $ 0.000000037$  & $ 0.964199$   & $1.8\times10^{-7}$ & $5 \times 10^{-6}$\\
      \hline
        $ 0.0001$   & $0.01$   & $ 0.000000072$  & $ 0.967174$   & $3.2\times10^{-7}$ & $\times 10^{-3}$\\
      \hline
        $ 0.0001$   & $0.05$   & $ 0.00000037$  & $ 0.964199$   & $1.4\times10^{-7}$ & $0.5$\\
      
       \hline
      \end{tabular}
      \end{center}
 \vspace{.25cm}

\subsubsection{Case-2: $\Gamma(T)=C_T T$}
Here we have considered $\Gamma= C_T T$ and thus we have $T=\frac{\Gamma}{C_T}$. The temperature of the thermal bath can be written using equation \eqref{105} as
 \begin{equation}
     T=\left(  \frac{3 r \dot{\chi}^2}{4C}\right)^{\frac{1}{4}}
     \label{202}
 \end{equation}
 and using equation \eqref{202}, the factor $T/H$ becomes
 \begin{equation}
    \frac{T}{H}=1.47 \left(\frac{r^{\frac{1}{4}} M_P^{\frac{7}{6}} \Lambda^{\frac{2}{3}} \left(\sin{\frac{\chi}{f}}\right)^{\frac{1}{6}}}{C^{\frac{1}{4}} f^{\frac{1}{6}} \mu^{\frac{5}{3}} \left(1+\cos{\frac{\chi}{f}}\right)^{\frac{7}{12}}}\right)
     \label{203} 
 \end{equation} 
 Equating $T=\frac{\Gamma}{C_T}$ with equation  \eqref{202}, we get
  
 \begin{equation}
 r=\left(\frac{0.0576 M_P^{\frac{14}{3}} \Lambda^{\frac{8}{3}} C_T^4 \left(\sin \left(\frac{\chi}{f}\right)\right)^{\frac{2}{3}}}{C f^{\frac{2}{3}}\mu^{\frac{20}{3}} \left( \cos \left(\frac{\chi}{f}\right)+1\right)^{\frac{7}{3}}}\right)^{\frac{1}{3}}   
\end{equation}
  
 The above equation is substituted in equation \eqref{121}, and $P_R$ is expressed in terms of $f$, $\mu$, $\Lambda$, $C_T$ and $\chi_i$. 
 
 \begin{equation}
    P_R= \frac{0.02827 f^{4/9} \mu^{40/9} C_T^{1/3} \left(\cos \left(\frac{\chi_i}{f}\right)+1\right)^{14/9}}{M_P^{28/9} C^{1/3} \Lambda^{16/9} \sin ^{\frac{4}{9}}\left(\frac{\chi_i}{f}\right)} 
    \label{201}
 \end{equation}
 
 The spectral index and tensor to scalar ratio at initial field value can be obtained using equations  \eqref{201}, \eqref{122}, \eqref{124}, \eqref{125}, and \eqref{127}. We carry out the numerical analysis for 60 e-foldings and find the range of model parameters for which the cosmological observables lie with the Planck bound. The possible values of the observables for different sets of model parameters are summarized in the following table 2.

\vspace{.25cm}
\begin{center}
    Table 2:   The values of the spectral index $n_s$ and tensor to scalar ratio $R$ for different values of model parameters $ f$, $\mu$ and $\Lambda$ (in units of $M_P$) and $C_T$ with $N=60$
\end{center}
  \begin{center}
      \begin{tabular}{|c|c|c|c|c|c|c|}
      \hline
      $\mu$   & $ f $  & $ \Lambda$   & $ n_s$   & $ R$ & $C_T$\\ 
      
      \hline
      $ 0.0001$   & $0.005$   & $ 0.000000038$  & $ 0.964822$   & $2.2\times10^{-7}$ & $0 .000005$\\
      \hline
        $ 0.0001$   & $0.01$   & $ 0.000000075$  & $ 0.966602$   & $2.07\times10^{-7}$ & $ 0.0001$\\
      \hline
        $ 0.0001$   & $0.05$   & $ 0.00000038$  & $ 0.964822$   & $2.7\times10^{-7}$ & $0.027$\\
      
       \hline
      \end{tabular}
      \end{center}
 \vspace{.25cm}
From the table, it is seen that the axion decay constant $f$ can be lowered below the GUT scale in this case. The lower bound on $C_T$ is obtained from the condition $T/H\geq1$ and the upper bound is given by the condition $r<1$. The spectral index in this case does not depend on the parameter $C_T$.

\subsubsection{Case-3: $\Gamma(\chi, T)=C_\chi \frac{T^3}{\chi^2} $}
 In the following, we find the expression of $r$ in this case also which will then be used to evaluate the perturbation spectrum. The temperature of the thermal bath can be written using equation \eqref{105} as
 \begin{equation}
     T=\left(  \frac{3 r \dot{\chi}^2}{4C}\right)^{\frac{1}{4}}
     \label{119}
 \end{equation}
 and using equation \eqref{128}, the factor $T/H$ becomes
 \begin{equation}
    \frac{T}{H}=1.47 \left(\frac{r^{\frac{1}{4}} M_P^{\frac{7}{6}} \Lambda^{\frac{2}{3}}\left(\sin{\frac{\chi}{f}}\right)^{\frac{1}{6}}}{C^{\frac{1}{4}} f^{\frac{1}{6}} \mu^{\frac{5}{3}} \left(1+\cos{\frac{\chi}{f}}\right)^{\frac{7}{12}}}\right)
     \label{126} 
 \end{equation}
 
 Equating equation \eqref{119} with $T=\left(\frac{3r\chi^2 H}{C_\chi}\right)^{\frac{1}{3}} $, the expression of $r$ can be obtained as
 \begin{equation}
 r=\frac{0.015625 M_P^6 \Lambda^8 C_\chi^4 \sin ^2\left(\frac{\chi}{f}\right)}{\mu^4 \chi^8 \left(f^{2/3} C \cos \left(\frac{\chi}{f}\right)+f^{2/3} C\right)^3}   
\end{equation}
 Substituting $r$ in equation \eqref{121}, the scalar power spectrum reads:
 
  \begin{equation}
    P_R= \frac{0.012678 \mu^4 C_\chi}{M_P^2  \chi^2 C} \left(\cos \left(\frac{\chi_i}{f}\right)+1\right) 
    \label{123}
 \end{equation}
 
 Using equations \eqref{123}, \eqref{122}, \eqref{124}, \eqref{125}, and \eqref{127} the expressions for the spectral index $n_s$ and tensor to scalar ratio $ R$ in terms of $f$, $\mu$, $\Lambda$ and $C_\chi$ are obtained for 60 e-folds. Since the expressions are lengthy, we do not present them here. We find the range of parameters where the spectral index lies within the PLANCK 2018 data. Similar to the previous case, here also the lower bound on $C_\chi$ is obtained from the condition $T/H\geq1$ and the upper bound is given by the condition $r<1$. It is found that the spectral index does not depend on the parameter $C_\chi$, while tensor to scalar ratio increases with decrease in $C_\chi$. The results that we obtain are presented in the table 3. From the table it is observed that consistent results are obtained at sub-Planckian scale.

       \vspace{.5cm}
\begin{center}
    Table 3:   The values of the spectral index $n_s$ and tensor to scalar ratio $R$ for different values of model parameters $ f$, $\mu$ and $\Lambda$ (in units of $M_P$) and $C_\chi$ with $N=60$
\end{center}
  \begin{center}
      \begin{tabular}{|c|c|c|c|c|c|c|}
      \hline
      $\mu$   & $ f $  & $ \Lambda$   & $ n_s$   & $ R$ & $C_\chi$\\ 
      \hline
        $ 0.0001$   & $0.01$   & $ 0.00000004$  & $ 0.963946$   & $1.35\times10^{-7}$ & $ 10^{7}$\\
      \hline
        $ 0.0001$   & $0.05$   & $ 0.0000002$  & $ 0.963098$   & $2.26\times10^{-7}$ & $1.5 \times 10^{8}$\\
      
       \hline
      \end{tabular}
      \end{center}
      \vspace{.25cm}

  \subsection{Strong Dissipative Regime}
   
  In strong dissipative regime, $r>>1$. So, the slow-roll parameters $\epsilon$, $\eta$ and $\beta$ in this regime become,
  \begin{equation}
    \epsilon=\frac{M_P^2}{2 f^2\left(\frac{\dot{\chi}^2}{\Lambda^4}+r\right)} \left(\frac{1-\cos{\frac{\chi}{f}}}{1+\cos{\frac{\chi}{f}}}\right) 
    \label{130}
  \end{equation}
  \begin{equation}
       \eta = -\frac{M_P^2}{f^2 \left(\frac{\dot{\chi}^2}{\Lambda^4}+r\right)}  \left(\frac{\cos{\frac{\chi}{f}}}{1+\cos{\frac{\chi}{f}}}\right) 
       \label{131}
  \end{equation}
  \begin{equation}
   \beta= \frac{2 M_P^2}{ f \left(\frac{\dot{\chi}^2}{\Lambda^4}+r\right)\dot{\chi}}  \left(\frac{\sin{\frac{\chi}{f}}}{1+\cos{\frac{\chi}{f}}}\right)   
   \label{132}
  \end{equation}
  Here also we used $\frac{\dot{\chi}^2}{2}>>\Lambda^4$ (slow-roll hierarchy scaling limit). And the equation of motion of the inflaton field reads:
 \begin{equation}
    3H\dot{\chi}\left(\frac{\dot{\chi}^2}{\Lambda^4}  +r\right)=\frac{\mu^4}{f}\sin\left(\frac{\chi}{f}\right)
    \label{133}
\end{equation}

 \subsubsection{Case-1: $\Gamma=\Gamma_0$}
  Solving equation \eqref{133} for $\dot{\chi}$ and substituting $\dot{\chi}$ in equations \eqref{130} and \eqref{131} as $\beta=0$, we get the slow-roll parameters $\epsilon$ and $\eta$ in terms of model parameters and $\chi$.  Inflation ends when either of the two parameters becomes of the order of unity. It is checked that both the parameters reach unity at the same time. So, we consider the equation $\epsilon=1$ to fix the final field value i.e. $\chi_f$. Initial field value can be obtained by using equation \eqref{115} and considering $N=60$. At this initial field value, the allowed values of cosmological observables are computed for different choices of model parameters. We present our results in the following table 4.

 \vspace{.5cm}
\begin{center}
    Table 4:   The values of the spectral index $n_s$ and tensor to scalar ratio $R$ for different values of model parameters $ f$, $\mu$ and $\Lambda$ (in units of $M_P$) and $r$ with $N=60$
\end{center}
  \begin{center}
      \begin{tabular}{|c|c|c|c|c|c|c|}
      \hline
      $\mu$   & $ f $  & $ \Lambda$   & $ n_s$   & $ R$ & $r$\\ 
      
      \hline
      $ 0.0001$   & $0.05$   & $ 0.00000037$  & $ 0.964456$   & $2.68\times10^{-7}$ & $ 1.1$\\
      \hline
        $ 0.0001$   & $0.1$   & $ 0.00000075$  & $ 0.96256$   & $4.08\times10^{-7}$ & $ 12.37$\\
       \hline
      \end{tabular}
      \end{center}
 \vspace{.25cm}
  
 \subsubsection{Case-2: $\Gamma(T)=C_T T$} 
 
 Equating $T=\frac{\Gamma}{C_T}$ with $ T=\left(  \frac{3 r \dot{\chi}^2}{4C}\right)^{\frac{1}{4}}$ (using equation \eqref{105}), we get
 \begin{equation}
     \dot{\chi}^6=  \left(\frac{108 C H^4 }{C_T^4}\right)r^3
     \label{135}
 \end{equation}
Solving equations \eqref{133} and \eqref{135}, the expressions of $\dot{\chi} $ and $r$ can be obtained. These expressions are substituted in equations \eqref{130}, \eqref{131} and \eqref{132} to obtain the slow-roll parameters $\epsilon$, $\eta$ and $\beta$ in terms of model parameters. In this case, we have numerically checked that inflation ends when the slow-roll conditions are no longer fulfilled i.e. when  $\epsilon$ and $\eta$ reach unity. This condition gives the final field value $(\chi_f)$. Now, using the definition of $N$ from equation \eqref{115}, a relation between $\chi_i$ and $\chi_f$ can be derived which is used to obtain the initial field value $(\chi_i)$. From the expressions of the power spectrum $P_R$ (equation \eqref{117}) and $P_T$ (equation \eqref{118}), the spectral index $n_s$ and tensor to scalar ratio $R$ can be computed at $\chi_i$ for different choice of parameters $f$, $\mu$, $\Lambda$ and $C_T$ for 60 e-foldings. In this case, both spectral index and tensor to scalar ratio depend on $C_T$. The possible values of the observables for different sets of model parameters are summarized in the following table 5. 

  \vspace{.5cm}
\begin{center}
    Table 5:   The values of the spectral index $n_s$ and tensor to scalar ratio $R$ for different values of model parameters $ f$, $\mu$ and $\Lambda$ (in units of $M_P$) and $C_T$ with $N=60$
\end{center}
  \begin{center}
      \begin{tabular}{|c|c|c|c|c|c|c|}
      \hline
      $\mu$   & $ f $  & $ \Lambda$   & $ n_s$   & $ R$ & $C_T$\\ 
      
      \hline
      $ 0.0001$   & $0.05$   & $ 0.00000038$  & $ 0.964738$   & $1.8\times10^{-7}$ & $ 0.09$\\
      \hline
        $ 0.0001$   & $0.1$   & $ 0.00000075$  & $ 0.967454$   & $2.5\times10^{-7}$ & $ 0.5$\\
       \hline
      \end{tabular}
      \end{center}
 \vspace{.25cm}

\subsubsection{Case-3: $\Gamma(\chi, T)=C_\chi \frac{T^3}{\chi^2} $}
 
 From $\Gamma(\chi, T)=C_\chi \frac{T^3}{\chi^2} $, we have $T=\left(\frac{3r\chi^2 H}{C_\chi}\right)^{\frac{1}{3}} $. Equating this with $ T=\left(  \frac{3 r \dot{\chi}^2}{4C}\right)^{\frac{1}{4}}$ (using equation \eqref{105}), we get
 \begin{equation}
     \dot{\chi}^6=3 r (4 C)^3 \left(\frac{\chi^2 H}{C_\chi}\right)^4
     \label{134}
 \end{equation}
 Expressions of $\dot{\chi}$ and $r$ can be obtained by solving equations \eqref{133} and \eqref{134}. Then following the similar procedure mentioned in the previous case, the spectral index $n_s$ and tensor to scalar ratio $R$ can be computed at $\chi_i$ for different model parameters $f$, $\mu$, $\Lambda$ and $C_\chi$ for 60 e-foldings. Here also both spectral index and tensor to scalar ratio depend on $C_\chi$. The possible values of the observables for different sets of model parameters are summarized in the following table 6. 
 
    \vspace{.5cm}
\begin{center}
    Table 6:   The values of the spectral index $n_s$ and tensor to scalar ratio $R$ for different values of model parameters $ f$, $\mu$ and $\Lambda$ (in units of $M_P$) and $C_\chi$ with $N=60$
\end{center}
  \begin{center}
      \begin{tabular}{|c|c|c|c|c|c|c|}
      \hline
      $\mu$   & $ f $  & $ \Lambda$   & $ n_s$   & $ R$ & $C_\chi$\\ 
      
      \hline
      $ 0.0001$   & $0.05$   & $ 0.0000002$  & $ 0.963135$   & $1.1\times10^{-7}$ & $ 3 \times10^8$\\
      \hline
        $ 0.0001$   & $0.1$   & $ 0.0000004$  & $ 0.963217$   & $1.3\times10^{-7}$ & $ 10^9$\\
       \hline
      \end{tabular}
      \end{center}
 \vspace{.25cm}
     
From the above six cases, we can conclude that a value of the axion decay constant of the order of GUT scale can be produced in this model for which the value of tensor to scalar ratio $R$ is predicted to be extremely low.

       \vspace{.5cm}
     \section{ Conclusions}
 
 The theory of cosmological inflation has drawn considerable attention in recent years due to its remarkable predictions about the early universe. In this work, we studied chromo-natural inflation in the context of warm inflation. The coupling between inflaton with other fields is included in warm chromo-natural inflation during inflationary phase which generates a temperature throughout inflation. So, the primordial power spectrum for warm chromo-natural inflation is different from that for chromo-natural inflation. 
   
 In this work, we analyzed warm chromo-natural inflation with constant $(\Gamma_0)$, temperature-dependent $(C_T T)$ and temperature-inflaton dependent $(C_\chi \frac{T^3}{\chi^2})$  dissipation coefficient. We considered both the weak (small $r$) and strong (large $r$) dissipative regimes and the analysis are carried out  separately in these two regimes. In each case, we parameterized the scalar and tensor power spectrum in terms of model parameters and obtained the parameter space  by taking $N=60$ and fixing $P_R=2\times10^{-9}$.
   
 When $\Gamma$ is considered to be a constant, the maximum and minimum values of $r$ are calculated for both the regime, where the $n_s$ and $R$ values lie within the observational limit. For linear dissipation coefficient $(\propto T)$ and cubic dissipation coefficient $(\propto T^3)$, we obtained the range of the parameter $C_\chi$ and $C_T$. The range of  $C_\chi$ and  $C_T$ in weak dissipative regime are $10^7\le C_\chi\le 1.5\times10^8$ and $0.000005\le C_T\le 0.027$, while $C_\chi$ and $C_T$ correspond to  higher values i.e. $3\times10^8\le C_\chi\le 10^9$ and $0.09\le C_T\le 0.5$ in the strong dissipative regime. In weak dissipative regime, the spectral index does not depend on $C_\chi$ and $C_T$, while in strong dissipative regime, both the observables, i.e. $n_s$ and $R$ depend on the parameters $C_\chi$ and $C_T$.
   
 This model allows to decrease the value of the axion decay constant $f$ from Planck scale in every cases and it is possible to lower the value of $f$ below the GUT scale in the weak dissipative regime. We calculated the spectral index $n_s$ and tensor-to-scalar ratio $R$ for different sets of parameters which are consistent with the current Planck data. This suggests that the warm chromo-natural inflation model is a viable inflationary model. The value of tensor to scalar ratio for both the regimes are obtained to be vanishingly small.

In this study, we consider the linear coupling of the scalar field with the Chern-Simon term. To study warm chromo-natural inflation further, one can take sinusoidal coupling of scalar field with the Chern-Simon term in the action.

  \printbibliography

@article{PhysRevD.23.347,
  title = {Inflationary universe: A possible solution to the horizon and flatness problems},
  author = {Guth, Alan H.},
  journal = {Phys. Rev. D},
  volume = {23},
  issue = {2},
  pages = {347--356},
  numpages = {0},
  year = {1981},
  month = {Jan},
  publisher = {American Physical Society}
}

@ARTICLE{1982PhLB..108..389L,
       author = {{Linde}, A.~D.},
        title = "{A new inflationary universe scenario: A possible solution of the horizon, flatness, homogeneity, isotropy and primordial monopole problems}",
      journal = {Physics Letters B},
         year = 1982,
        month = feb,
       volume = {108},
       number = {6},
        pages = {389-393}
}

@article{LINDE1983177,
title = {Chaotic inflation},
journal = {Physics Letters B},
volume = {129},
number = {3},
pages = {177-181},
year = {1983},
author = {A.D. Linde}
}

@misc{linde2005particle,
      title={Particle Physics and Inflationary Cosmology}, 
      author={Andrei Linde},
      year={2005},
      eprint={hep-th/0503203},
      archivePrefix={arXiv},
      primaryClass={hep-th}
}

@article{PhysRevD.49.748,
  title = {Hybrid inflation},
  author = {Linde, Andrei},
  journal = {Phys. Rev. D},
  volume = {49},
  issue = {2},
  pages = {748--754},
  numpages = {0},
  year = {1994},
  month = {Jan},
  publisher = {American Physical Society}
}

@article{PhysRevLett.48.1220,
  title = {Cosmology for Grand Unified Theories with Radiatively Induced Symmetry Breaking},
  author = {Albrecht, Andreas and Steinhardt, Paul J.},
  journal = {Phys. Rev. Lett.},
  volume = {48},
  issue = {17},
  pages = {1220--1223},
  numpages = {0},
  year = {1982},
  month = {Apr},
  publisher = {American Physical Society}
}

@article{PhysRevLett.75.3218,
  title = {Warm Inflation},
  author = {Berera, Arjun},
  journal = {Phys. Rev. Lett.},
  volume = {75},
  issue = {18},
  pages = {3218--3221},
  numpages = {0},
  year = {1995},
  month = {Oct},
  publisher = {American Physical Society} 
}

@article{PhysRevD.55.3346,
  title = {Interpolating the stage of exponential expansion in the early universe: Possible alternative with no reheating},
  author = {Berera, Arjun},
  journal = {Phys. Rev. D},
  volume = {55},
  issue = {6},
  pages = {3346--3357},
  numpages = {0},
  year = {1997},
  month = {Mar},
  publisher = {American Physical Society}
}

@article{54db6c2c12e0450d9e603224eb51880c,
title = "Warm inflation in the adiabatic regime - A model, an existence proof for inflationary dynamics in quantum field theory",
author = "Arjun Berera",
year = "2000",
month = oct,
day = "9",
language = "English",
volume = "585",
pages = "666--714",
journal = "Nuclear physics b",
publisher = "Elsevier",
number = "3",
}

@article{PhysRevD.58.123508,
  title = {Strong dissipative behavior in quantum field theory},
  author = {Berera, Arjun and Gleiser, Marcelo and Ramos, Rudnei O.},
  journal = {Phys. Rev. D},
  volume = {58},
  issue = {12},
  pages = {123508},
  numpages = {20},
  year = {1998},
  month = {Nov},
  publisher = {American Physical Society}
}

@article{PhysRevLett.65.3233,
  title = {Natural inflation with pseudo Nambu-Goldstone bosons},
  author = {Freese, Katherine and Frieman, Joshua A. and Olinto, Angela V.},
  journal = {Phys. Rev. Lett.},
  volume = {65},
  issue = {26},
  pages = {3233--3236},
  numpages = {0},
  year = {1990},
  month = {Dec},
  publisher = {American Physical Society} 
}

@article{Adams_1993,
	year = 1993,
	month = {jan},
  	publisher = {American Physical Society ({APS})},
    volume = {47},
  
	number = {2},
  
	pages = {426--455},
  
	author = {Fred C. Adams and J. Richard Bond and Katherine Freese and Joshua A. Frieman and Angela V. Olinto},
  
	title = {Natural inflation: Particle physics models, power-law spectra for large-scale structure, and constraints from the Cosmic Background Explorer},
  
	journal = {Physical Review D}
}

@article{Savage_2006,
 
 	year = 2006,
	month = {dec},
  
	publisher = {American Physical Society ({APS})},
  
	volume = {74},
  
	number = {12},
  
	author = {Christopher Savage and Katherine Freese and William H. Kinney},
  
	title = {Natural inflation: Status after {WMAP} 3-year data},
  
	journal = {Physical Review D}
}

@article{2015,
   title={Natural inflation: consistency with cosmic microwave background observations of Planck and BICEP2},
   volume={2015},
   journal={Journal of Cosmology and Astroparticle Physics},
   publisher={IOP Publishing},
   author={Freese, Katherine and Kinney, William H.},
   year={2015},
   month={Mar},
   pages={044–044}
}

@article{2004,
   title={On natural inflation},
   volume={70},
   journal={Physical Review D},
   publisher={American Physical Society (APS)},
   author={Freese, Katherine and Kinney, William H.},
   year={2004},
   month={Oct}
}

@article{Banks_2003,
	 year = 2003,
	month = {jun},
  
	publisher = {{IOP} Publishing},
  
	volume = {2003},
  
	number = {06},
  
	pages = {001--001},
  
	author = {Tom Banks and Michael Dine and P J Fox and E Gorbatov},
  
	title = {On the possibility of large axion decay constants},
  
	journal = {Journal of Cosmology and Astroparticle Physics}
}

@article{2012,
   title={Natural Inflation on a Steep Potential with Classical Non-Abelian Gauge Fields},
   volume={108},
   journal={Physical Review Letters},
   publisher={American Physical Society (APS)},
   author={Adshead, Peter and Wyman, Mark},
   year={2012},
   month={Jun}
}

@article{PhysRevD.87.103501,
  title = {Stability analysis of chromo-natural inflation and possible evasion of Lyth's bound},
  author = {Dimastrogiovanni, Emanuela and Peloso, Marco},
  journal = {Phys. Rev. D},
  volume = {87},
  issue = {10},
  pages = {103501},
  numpages = {19},
  year = {2013},
  month = {May},
  publisher = {American Physical Society}
}

@article{2016,
   title={Consistency of warm 
k
-inflation},
   volume={94},
   number={10},
   journal={Physical Review D},
   publisher={American Physical Society (APS)},
   author={Peng, Zhi-Peng and Yu, Jia-Ning and Zhu, Jian-Yang and Zhang, Xiao-Min},
   year={2016},
   month={Nov}
}

@article{2018,
   title={Perturbation spectra in the warm 
k
-inflation},
   volume={97},
   journal={Physical Review D},
   publisher={American Physical Society (APS)},
   author={Peng, Zhi-Peng and Yu, Jia-Ning and Zhang, Xiao-Min and Zhu, Jian-Yang},
   year={2018},
   month={Mar}
}

@article{Seery_2005,
	year = 2005,
	month = {jun},
	publisher = {{IOP} Publishing},
	volume = {2005},
	number = {06},
	pages = {003--003},
	author = {David Seery and James E Lidsey},
	title = {Primordial non-Gaussianities in single-field inflation},
	journal = {Journal of Cosmology and Astroparticle Physics}
}

@article{Berera_2009,
 
	year = 2009,
	month = {jan},
  
	publisher = {{IOP} Publishing},
  
	volume = {72},
  
	number = {2},
  
	pages = {026901},
  
	author = {Arjun Berera and Ian G Moss and Rudnei O Ramos},
  
	title = {Warm inflation and its microphysical basis},
  
	journal = {Reports on Progress in Physics}
}

@article{PhysRevD.62.083517,
  title = {Perturbation spectra in the warm inflationary scenario},
  author = {Taylor, A. N. and Berera, A.},
  journal = {Phys. Rev. D},
  volume = {62},
  issue = {8},
  pages = {083517},
  numpages = {11},
  year = {2000},
  month = {Sep},
  publisher = {American Physical Society},
  
}

@article{Adshead_2013,
	 
  
	year = 2013,
	month = {sep},
  
	publisher = {Springer Science and Business Media {LLC}
},
  
	volume = {2013},
  
	number = {9},
  
	author = {Peter Adshead and Emil Martinec and Mark Wyman},
  
	title = {Perturbations in Chromo-Natural Inflation},
  
	journal = {Journal of High Energy Physics}
}

@article{Dimastrogiovanni_2013,
	 year = 2013,
	month = {feb},
  
	publisher = {{IOP} Publishing},
  
	volume = {2013},
  
	number = {02},
  
	pages = {046--046},
  
	author = {Emanuela Dimastrogiovanni and Matteo Fasiello and Andrew J Tolley},
  
	title = {Low-energy effective field theory for chromo-natural inflation},
  
	journal = {Journal of Cosmology and Astroparticle Physics}
}

@article{Bhattacharjee_2015,
	 year = 2015,
	month = {apr},
  
	publisher = {World Scientific Pub Co Pte Lt},
  
	volume = {30},
  
	number = {11},
  
	pages = {1550040},
  
	author = {Anindita Bhattacharjee and Atri Deshamukhya and Sudhakar Panda},
  
	title = {A note on low energy effective theory of chromo-natural inflation in the light of {BICEP}2 results},
  
	journal = {Modern Physics Letters A}
}

@article{DESHAMUKHYA_2009,
	 year = 2009,
	month = {dec},
  
	publisher = {World Scientific Pub Co Pte Lt},
  
	volume = {18},
  
	number = {14},
  
	pages = {2093--2106},
  
	author = {Atri Deshamukhya and Sudhakar Panda},
  
	title = {{Warm} {tachyonic} {inflation} {in} a {warped} {background}
},
  
	journal = {International Journal of Modern Physics D}
}

@article{Zhang_2009,
	 year = 2009,
	month = {mar},
  
	publisher = {{IOP} Publishing},
  
	volume = {2009},
  
	number = {03},
  
	pages = {023--023},
  
	author = {Yi Zhang},
  
	title = {Warm inflation with a general form of the dissipative coefficient},
  
	journal = {Journal of Cosmology and Astroparticle Physics}
}

@article{Hall_2004,
	 year = 2004,
	month = {apr},
  
	publisher = {American Physical Society ({APS})},
  
	volume = {69},
  
	number = {8},
  
	author = {Lisa M. H. Hall and Ian G. Moss and Arjun Berera},
  
	title = {Scalar perturbation spectra from warm inflation},
  
	journal = {Physical Review D}
}

@article{Moss_2008,
	 
  
	 
  
	year = 2008,
	month = {nov},
  
	publisher = {{IOP} Publishing},
  
	volume = {2008},
  
	number = {11},
  
	pages = {023},
  
	author = {Ian G Moss and Chun Xiong},
  
	title = {On the consistency of warm inflation},
  
	journal = {Journal of Cosmology and Astroparticle Physics}
}

@article{Bastero_Gil_2005,
	 
  
	year = 2005,
	month = {mar},
  
	publisher = {American Physical Society ({APS})},
  
	volume = {71},
  
	number = {6},
  
	author = {Mar Bastero-Gil and Arjun Berera},
  
	title = {Determining the regimes of cold and warm inflation in the supersymmetric hybrid model},
  
	journal = {Physical Review D}
}
 \end{document}